\documentclass[pra,twocolumn,superscriptaddress]{revtex4}

\usepackage{amsmath}
\usepackage{latexsym}
\usepackage{amssymb}
\usepackage{graphicx}
\usepackage[colorlinks=true, citecolor=blue, urlcolor=blue]{hyperref}
\usepackage{float}
\usepackage{amsfonts}
\usepackage{textcomp}
\usepackage{mathpazo}
\usepackage{comment}
\usepackage{amsmath,bbm}

\newcommand{\tr}{{\rm Tr}}
\newcommand{\idol}{\ensuremath{\mathbbm 1}}

\begin{document}
\title{Detecting coherence with respect to general quantum measurements}
\author{Yu-Cheng Chen}
\affiliation{School of Physical Science and Technology, Ningbo University, Ningbo, 315211, China}
\author{Jiong Cheng}
\affiliation{School of Physical Science and Technology, Ningbo University, Ningbo, 315211, China}
\author{Wen-Zhao Zhang}
\email{zhangwenzhao@nbu.edu.cn}
\affiliation{School of Physical Science and Technology, Ningbo University, Ningbo, 315211, China}
\author{Cheng-Jie Zhang}
\email{chengjie.zhang@gmail.com}
\affiliation{School of Physical Science and Technology, Ningbo University, Ningbo, 315211, China}
\affiliation{State Key Laboratory of Precision Spectroscopy, School of Physics and Electronic Science, East China Normal University, Shanghai 200241, China}

\begin{abstract}
Quantum coherence is a crucial resource in quantum resource theory. Previous study mainly focused on standard coherence under a complete orthogonal reference basis. The standard coherence has recently been extended to general positive-operator-valued measure (POVM)-based coherence, including block coherence as a special case. Therefore, it is necessary to construct block coherence and POVM-based coherence witnesses to detect them. In this work, we present witnesses for block coherence and POVM-based coherence, and obtain the necessary and sufficient conditions for arbitrary block coherence and POVM-based coherence witnesses. We also discuss possible realizations of some block coherence and POVM-based coherence witnesses in experiments, and present examples of measuring block coherence witnesses based on real experimental data. Furthermore, an application of block coherence witnesses has been presented in a quantum parameter estimation task with a degenerate Hamiltonian, and one can estimate the unknown parameter by measuring our block coherence witnesses if the input state is block coherent. Lase but not least, we prove that the quantum Fisher information of any block incoherent state is equal to zero, which coincides with the result from measuring block coherence witnesses.
\end{abstract}
\date{\today}

\maketitle

\section{Introduction}
Quantum coherence plays a significant role in quantum theory, which has a number of applications in quantum optics, quantum information processing, nanoscale thermodynamics and biological systems \cite{MMTC,SMP,FGM,FGSL,MJJ,YYZW,MBS,DZLX,KDWZ,ZTYH,DGGG,TMG,CLYG,KPF,MKT}. Recently, it has been recognized that coherence can be treated as one kind of quantum resources. Therefore, detecting coherence is crucial in quantum physics.

In Ref. \cite{TGM}, quantitative investigations of quantum coherence have been launched, and several coherence measures for standard coherence, with respect to von Neumann measurements, have been proposed. For standard coherence, consider a $d$-dimensional Hilbert space $\mathcal{H}$, a state $\delta$ is incoherent under a chosen reference basis $\{|i\rangle\}_{i=1}^{d}$ if and only if $\delta$ is diagonal under the reference basis \cite{TGM}, i.e.,
\begin{equation}\label{}
  \delta =\sum_{i=1}^{d} p_i|i\rangle\langle i|,
\end{equation}
with probabilities $\{p_i\}$. One can define the standard dephasing operation $\Delta$ as
\begin{equation}\label{de}
\Delta(\rho):=\sum_{i=1}^{d}|i\rangle\langle i|\rho|i\rangle\langle i|.
\end{equation}
Thus, a state  $\delta$ is incoherent under a chosen reference basis $\{|i\rangle\}_{i=1}^{d}$ if and only if the following condition holds,
\begin{equation}\label{}
  \delta=\Delta(\delta),
\end{equation}
where $\Delta$ is defined in Eq. (\ref{de}).


In Ref. \cite{Aberg}, {\AA}berg proposed a framework which actually defined block coherence with respect to projective measurements. The standard coherence can be viewed as a special case of block coherence. Moreover, Bischof \textit{et al.} generalized block coherence to coherence with respect to the most general quantum measurements \cite{FHD}, i.e., the positive-operator-valued measure (POVM)-based coherence. Therefore, the resource theory of coherence was generalized by extending the standard coherence to  block coherence and even  POVM-based coherence. It is worth noticing that POVMs describe the most general type of quantum measurements, and they may be more advantageous compared to von Neumann measurements.

However, unlike the standard coherence, there are only a few results reported on quantifying block coherence and POVM-based coherence \cite{FHD,CMWR,QCWRT}. The methods to detect whether a state has nonzero block coherence and POVM-based coherence are missing. For the standard coherence, Ref. \cite{CTRM} first introduced the standard coherence witness $W$. Similar to entanglement witnesses, a Hermitian operator $W$ is a standard coherence witness if $\tr(\delta W)\geq0$ holds for all incoherent states $\delta$. If one finds $\tr(\rho W)<0$ for a state $\rho$, then the state $\rho$ must be a standard coherent state. Compared with coherence measures, which usually need full information of the state by using quantum state tomography, coherence witnesses can be measured with much less measurements since quantum state tomography requires exponentially growing measurements with the number of qubits.
Thus, it is necessary to construct block coherence witnesses and POVM-based coherence witnesses to detect them without quantum state tomography, especially for experimentally unknown states.

The purpose of this work is two-fold. On the one hand,  we present witnesses for block coherence and POVM-based coherence, and obtain the necessary and sufficient conditions for arbitrary block coherence witnesses and POVM-based coherence witnesses. Moreover, we discuss possible realizations of some block coherence and POVM-based coherence witnesses in real experiments, and present  examples to detect block coherence coherence by measuring block coherence witnesses. On the other hand, we provide an application of block coherence witnesses in a quantum parameter estimation task with a degenerate Hamiltonian, and one can estimate the unknown parameter by measuring our block coherence witnesses if the input state is block coherent. We also prove that the quantum Fisher information of any block incoherent state is equal to zero, which coincides with the result from measuring block coherence witnesses.

\section{Detecting block coherence based on block coherence witnesses}
Before embarking on our main results, let us first review the definition of block incoherent states. 
In Refs. \cite{FHD,CMWR,QCWRT,CCBP,Aberg,WJD}, block incoherent state has been defined as follows.

Given a $d$-dimensional Hilbert space $\mathcal{H}$, which has been divided into $n$ ($n\leq d$) subspaces, and subspace projectors are $\mathbf{P}:=\{P_s\}_{s=1}^{n}$ with $\sum_{s=1}^n P_s=\idol$ (where $\idol$ is the identity operator). A state $\widetilde{\delta}$ is block incoherent under the reference subspace projectors $\mathbf{P}$, if and only if $\widetilde{\delta}$ is block diagonal under the reference $\mathbf{P}$, i.e,
\begin{equation}\label{}
\widetilde{\delta}=\sum_{s=1}^{n}P_{s}\widetilde{\delta} P_{s}:=\widetilde{\Delta}(\widetilde{\delta}),
\end{equation}
where we define the modified dephasing operation as
\begin{equation}
\widetilde{\Delta}(\rho) := \sum_{s} P_s \rho P_s.
\end{equation}
Similar to the witnesses for standard coherence \cite{CTRM,ZZY}, we can construct block coherence witnesses as follows.

\textbf{Theorem 1.} (a) For any Hermitian operator $A$, we can construct a block coherence witness
\begin{equation}\label{10}
\widetilde{W}_A=\widetilde{\Delta}(A)- A.
\end{equation}
(b) An arbitrary Hermitian operator $\widetilde{W}$ is a block coherence witness if and only if $\widetilde{\Delta}(\widetilde{W})\geq 0$.

\textbf{Proof.---} (a) We first prove that $\widetilde{W}_A$ is a block coherence witness. Since $A$ is a Hermitian operator, $\widetilde{W}_A$ must be Hermitian as well. Thus, for an arbitrary block incoherent state $\widetilde{\delta}=\sum_{s}P_{s}\widetilde{\delta} P_{s}$, we can obtain
\begin{eqnarray}
\mathrm{Tr}(\widetilde{\delta} \widetilde{W}_A)  &=& \mathrm{Tr}[\widetilde{\delta}\widetilde{\Delta}(A)]-\tr[\widetilde{\delta}A] \nonumber\\
              &=& \mathrm{Tr}[\widetilde{\delta}\sum_{s} P_{s}A P_{s}] -\tr[\widetilde{\delta}A ] \nonumber\\
              &=& \mathrm{Tr}[\sum_{s}P_{s}\widetilde{\delta}P_{s}A] - \tr[\widetilde{\delta}A] \nonumber\\
                   &=& 0,
\end{eqnarray}
which means $\widetilde{W}_A$ is a block coherence witness.

(b) It is worth noticing that a Hermitian operator $W$ is a coherence witness  for standard coherence if and only if $\Delta(W) \geq 0$ \cite{CTRM}. Similarly, we can prove that a Hermitian operator $\widetilde{W}$ is a block coherence witness if and only if $\widetilde{\Delta}(\widetilde{W})\geq 0$.

Firstly, if $\widetilde\Delta(\widetilde{W})\geq 0$ holds, then for any block incoherent state $\widetilde{\delta}$ we can obtain that
\begin{eqnarray}
\tr[\widetilde{\delta} \widetilde{W}] &=&\tr[\widetilde\Delta(\widetilde{\delta}) \widetilde{W}]\nonumber \\
&=&\tr[\widetilde\Delta(\widetilde{W}) \widetilde{\delta}]\nonumber \\
& \geq& 0,
\end{eqnarray}
i.e., $\widetilde{W}$ is a block coherence witness.

Conversely, we prove that if for any block incoherent state $\widetilde{\delta}$ $\tr[\widetilde{\delta} \widetilde{W}]\geq 0$ holds, then $\widetilde\Delta(\widetilde{W})\geq 0$. For any quantum state $\rho$, one can obtain
\begin{eqnarray}
\tr[\rho\widetilde{\Delta}( \widetilde{W})]
 &=&\tr[\widetilde\Delta(\rho) \widetilde{W}] \nonumber \\
&=&\tr[\widetilde{\delta}_{\rho}\widetilde{W}] \nonumber \\
& \geq& 0,
\end{eqnarray}
where $\widetilde{\delta}_{\rho}:=\widetilde\Delta(\rho)$ is a block incoherent state.
Thus, $ \widetilde{\Delta}(\widetilde{W})$ is positive-semidefinite, i.e., $\widetilde\Delta(\widetilde{W})\geq 0$.

Therefore, it can be given that $\widetilde{W}$ is block coherence witness if and only if $\widetilde{\Delta}(\widetilde{W})\geq 0$. \hfill  $\square$

\textbf{Remark.---}  Based on Theorem 1(a), one can construct a block coherence witness $\widetilde{W}_{\sigma}$ by using any density matrix $\sigma$ as the Hermitian operator $A$ in Eq. (\ref{10}),
\begin{equation}\label{15}
\widetilde{W}_{\sigma} = \widetilde{\Delta}(\sigma) - \sigma.
\end{equation}
Moreover, if $\sigma$ is a pure state $|\phi\rangle$, we can obtain that
\begin{equation}\label{pureW1}
\widetilde{W}_{\phi} = \widetilde{\Delta}(|\phi\rangle\langle\phi|) -|\phi\rangle\langle\phi|,
\end{equation}
and
%
\begin{eqnarray}
\mathrm{Tr}[\rho \widetilde{W}_{\phi}]&=&\mathrm{Tr}[\rho(\widetilde{\Delta}(|\phi\rangle\langle\phi|) -|\phi\rangle\langle\phi|)]\nonumber \\
&=&\langle \phi|\widetilde\Delta(\rho)|\phi\rangle-\langle \phi|\rho|\phi\rangle  \nonumber\\
&=&F(\widetilde\Delta(\rho),|\phi\rangle)-F(\rho,|\phi\rangle),
\end{eqnarray}
where $F(\rho,|\phi\rangle):=\langle \phi|\rho|\phi\rangle$ is the fidelity between the state $\rho$ and the pure state $|\phi\rangle$. Therefore, the expect value of block coherence witness $ \widetilde{W}_{\phi}$ is related to the two fidelities.

\section{Coherence witness with respect to general measurements}
Recently, Bishof $et \ al.$ introduced a generalization to a resource theory of coherence with respect to the general quantum measurement. The POVM-based coherence is defined as follows \cite{FHD,CMWR,QCWRT}.

Let $\mathbf{E}$ be a  collection of $n$ positive operators $\mathbf{E}:=\{E_{i}\}_{i=1}^{n}$ with $\sum_{i=1}^{n} E_{i}=\idol$. The corresponding measurement operator of each $E_i$ is denoted by $A_i$, such that $E_{i}=A_{i}^{\dagger}A_{i}$ holds. Thus, a state $\bar{\delta}$ is called incoherent state with respect to the general measurement $\mathbf{E}$ if and only if
\begin{equation}
E_{i}\bar{\delta} E_{i'}=0, \ \forall i\neq i'.
\end{equation}
Note that it is equivalent to \cite{FHD}
\begin{equation}
A_{i}\bar{\delta} A_{i'}^{\dagger}=0, \ \forall i\neq i'.
\end{equation}
Therefore, any POVM-based incoherent state $\bar{\delta}$ should satisfy
\begin{equation}\label{bardelta}
\bar{\delta}= \sum_{i}E_{i}\bar{\delta} E_{i}:=\bar{\Delta}(\bar{\delta}),
\end{equation}
where $\bar{\Delta}$ is defined as
\begin{equation}
\bar{\Delta}(\rho) := \sum_{i} E_i \rho E_i.
\end{equation}
It is worth noticing that Eq. (\ref{bardelta}) can be easily  proved from the definition of POVM-based incoherent state, since for any POVM-based incoherent state $\bar{\delta}$, one can obtain
\begin{eqnarray}
\bar{\delta} &=&(\sum_{i}E_{i})\bar{\delta} (\sum_{j}E_{j}) \nonumber \\
&=&\sum_{i} E_{i}\bar{\delta} E_{i} +\sum_{i\neq j} E_{i}\bar{\delta} E_{j}\nonumber \\
&=&\sum_{i}E_{i}\bar{\delta} E_{i},
\end{eqnarray}
where we have used $E_{i}\bar{\delta} E_{j}=0, \forall i\neq j$.

\textbf{Theorem 2.} (a) For any Hermitian operator $A$, we can construct a POVM-based coherence witness $\bar{W}_A$  as follows,
\begin{equation}\label{}
\bar{W}_A=\bar{\Delta}  (A)- A.
\end{equation}
(b) An arbitrary Hermitian operator $\bar{W}$ is a POVM-based coherence witness if and only if $\bar{\Delta}(\bar{W})\geq 0$.

\textbf{Proof.---} (a) We provide that $\bar{W}_A$ is a POVM-based coherence witness. As $A$ is a Hermitian operator, $\bar{W}_A$ must be
Hermitian. For any incoherent state $\bar{\delta}$ with respect to $\{E_{i}\}$, $\bar{\delta}=\sum_{i}E_{i}\bar{\delta} E_{i}$ holds and thus
\begin{eqnarray}
\mathrm{Tr}(\bar{\delta} \bar{W}_A)  &=& \mathrm{Tr}[\bar{\delta}\bar{\Delta} (A)]-\tr[\bar{\delta} A] \nonumber \\
              &=& \mathrm{Tr}[\bar{\delta}\sum_{i} E_{i}A E_{i}] - \tr[\bar{\delta} A ] \nonumber \\
              &=& \mathrm{Tr}[\sum_{i}E_{i}\bar{\delta} E_{i}A] - \tr[\bar{\delta} A] \nonumber \\
                   &=& 0,
\end{eqnarray}
which means $\bar{W}_A$ is a POVM-based coherence witness.

(b) Firstly, if $\bar{\Delta}(\bar{W})\geq 0$ holds, then for  any POVM-based incoherent state $\bar{\delta}$ we can obtain that
\begin{eqnarray}
\tr[\bar{\delta} \bar{W}] &=&\tr[\bar{\Delta} (\bar{\delta}) \bar{W}]  \nonumber \\
&=&\tr[\bar{\Delta}(\bar{W}) \bar{\delta}] \nonumber \\
& \geq& 0.
\end{eqnarray}
Thus, $\bar{W}$ is a POVM-based  coherence witness.

Conversely, we prove that if $\tr[\bar{\delta} \bar{W}]\geq 0$, then $\bar{\Delta}(\bar{W})\geq 0$. For any quantum state $\rho$, one can obtain that
\begin{eqnarray}
\tr[\rho\bar{\Delta}(\bar{ W})]
 &=&\tr[\bar{\Delta}(\rho) \bar{W}] \nonumber \\
&=&\tr[\bar{\delta}_{\rho}\bar{W}] \nonumber \\
& \geq& 0,
\end{eqnarray}
where $\bar{\delta}_{\rho}:=\bar{\Delta}(\rho)$  is a POVM-based  incoherent state. Thus, $\bar{\Delta}(\bar{W})$ is positive-semidefinite, i.e., $\bar{\Delta}(\bar{W})\geq 0$.

Therefore, we proved that $\bar{W}$ is a POVM-based coherence witness if and only if $\bar{\Delta}(\bar{W})\geq 0$. \hfill  $\square$

\textbf{Remark.---} Based on Theorem 2(a), we can also construct a POVM-based coherence witness $\bar{W}_{\sigma}$ by choosing any density matrix $\sigma$ as the Hermitian operator $A$,
\begin{equation}\label{43}
\bar{W}_{\sigma} = \bar{\Delta}({\sigma}) - \sigma.
\end{equation}
Moreover, if $\sigma$ is a  pure state $|\phi\rangle$, one can obtain that
\begin{equation}\label{pureW2}
\bar{W}_{\phi} = \bar{\Delta}(|\phi\rangle\langle\phi|) - |\phi\rangle\langle\phi|,
\end{equation}
and
\begin{eqnarray}
\mathrm{Tr}[\rho \bar{W}_{\phi}]&=&\mathrm{Tr}[\rho(\bar{\Delta}(|\phi\rangle\langle\phi|) - |\phi\rangle\langle\phi|)] \nonumber\\
&=&\tr[\rho \bar{\Delta}(|\phi\rangle\langle\phi|)] -\tr[\rho |\phi\rangle\langle\phi|] \nonumber\\
&=&F( \bar{\Delta}(\rho), |\phi\rangle)-F(\rho,|\phi\rangle),
\end{eqnarray}
where it demonstrates the relationship of the POVM-based coherence witness $\bar{W}_{\phi}$ and fidelities between the state $\rho$ (or $\bar\Delta(\rho)$) and the pure state $|\phi\rangle$.

\section{Possible experimental realization for witnesses and examples}
Many experiments have measured fidelities $F=\langle\phi|\rho_{exp}|\phi\rangle$ between the experimental state $\rho_{exp}$ and a target pure state $|\phi\rangle$ \cite{ion,Nph1,Nph2,Nph3}. For bipartite and multipartite systems, one possible way to measure fidelities is to decompose the operator $|\phi\rangle\langle \phi|$ as sum of tensor products of local observables \cite{Nph1,Nph2,Nph3}. Therefore, one can measure our witnesses (\ref{pureW1}) and (\ref{pureW2}) in the same manner.

In the following, we will present examples of $N$-qubit W states $|W_N\rangle$ from real experimental data, where
\begin{equation}\label{W_N}
  |W_N\rangle=(|0\cdots 001\rangle+|0\cdots 010\rangle+ \cdots +|10\cdots0\rangle)/\sqrt{N}.
\end{equation}
We use the block coherence witness (\ref{pureW1}) to detect block coherence of W states. In Ref. \cite{ion}, $N$-qubit W states ($4\leq N\leq8$) have be experimentally generated by trapped ions, with fidelities between experimental states and perfect W states being $F_4=0.846$, $F_5=0.759$, $F_6=0.788$, $F_7=0.763$, $F_8=0.722$ for the $4$-, $5$-, $6$-, $7$- and $8$-ion W states, respectively. Moreover, numerical values of the density matrices of experiment states with $4\leq N\leq8$ have been presented in Ref. \cite{ion}. It is worth noticing that the experimental states have local phases, and one can find the local phases by maximizing the fidelities $F=\langle \tilde{W}_N|\rho_{exp}|\tilde{W}_N\rangle$ with $|\tilde{W}_N\rangle$ being W states containing local phases shown in \cite{ion}. After choosing local unitary transformations based on local phases, we can transform $\rho_{exp}$ to $\rho'_{exp}$ such that $F=\langle \tilde{W}_N|\rho_{exp}|\tilde{W}_N\rangle=\langle W_N|\rho'_{exp}|W_N\rangle$. We consider the following  reference subspace projectors $\mathrm{\mathbf{P}}$  with
\begin{eqnarray}
P_0&=&|\phi^-\rangle\langle \phi^-|,\nonumber\\
 P_1&=&|\phi^+\rangle\langle \phi^+|,\nonumber\\
 P_2&=&|0\cdots 010\rangle\langle 0\cdots010|,\nonumber\\
 &&\cdots,\nonumber\\
 P_{N-1}&=&|010\cdots0\rangle\langle 010\cdots0|,\nonumber\\
 P_N&=&\idol-\sum_{i=0}^{N-1}P_{i},\nonumber
\end{eqnarray}
where $|\phi^{\pm}\rangle:=(|00\cdots 01\rangle\pm |10\cdots0\rangle)/\sqrt{2}$. Thus, our block coherence witness is as follows,
\begin{eqnarray}
\widetilde{W}&=&\widetilde{\Delta}(|W_N\rangle\langle W_N|)- |W_N\rangle\langle W_N|\nonumber\\
&=&\sum_{i=0}^N P_i |W_N\rangle\langle W_N| P_i -|W_N\rangle\langle W_N|.
\end{eqnarray}
From Table \ref{lab1}, we can see that  $-\tr(\rho'_{exp} \widetilde{W})$ is always greater than zero, which means all the experimental states in Ref. \cite{ion} contain the block coherence under the above reference subspace projectors $\mathrm{\mathbf{P}}$.

\begin{table}
\begin{tabular}{c|ccccc}
\hline
\hline
  $|W_N\rangle$ & $4$ & $5$ & $6$ & $7$& $8$  \\
  \hline
  $F_{N}$  & $ 0.846 $ \ \ & $ 0.759 $ \ \  & $ 0.788 $\ \  & $ 0.763 $\ \  & $ 0.722 $ \\
  \hline
    $\langle \widetilde{\Delta}(|W_N\rangle\langle W_N|) \rangle$  & $0.321$ \ \ & $0.207$\ \ & $0.173$\ \ & $0.141$\ \  & $0.119$ \\
  \hline
  $-\tr(\rho'_{exp} \widetilde{W}) $ &0.525 \ \ &0.552 \ \ &0.615\ \  &0.622\ \ &0.603  \\
  \hline
  \hline
\end{tabular}
\caption{The fidelity $F_{N}$ is the result of overlap $\langle W_N |\rho'_{exp}|W_N\rangle$ from Ref. \cite{ion}. The values $\langle \widetilde{\Delta}(|W_N\rangle\langle W_N|) \rangle$ and $-\tr(\rho'_{exp} \widetilde{W})$ are obtained from the density matrices of experiment states in Ref. \cite{ion}. One can see that $-\tr(\rho'_{exp} \widetilde{W})$ is always greater than zero.  }\label{lab1}
\end{table}

\section{Quantum parameter estimation task with degenerate Hamiltionians}
For quantum metrology, one of the main tasks is  parameter estimation in a quantum channel to improve accuracy that is different from the standard quantum limits \cite{VGS,JPD}. Quantum coherence plays a fundamental role in quantum parameter estimation \cite{GTI}. For the case of unitary evolution with a degenerate Hamiltonian, we will propose a simple application of quantum block coherence, and find that the quantum Fisher information is strongly related to the block coherence.

\subsection{Quantum parameter estimation by using block coherent states}
Now we consider a $d$-dimensional Hilbert space, suppose that $H$ is a degenerate Hamiltonian,
\begin{equation}\label{1}
H=\sum_{s=1}^{n}\sum_{g=1}^{k_s}E_s|s,g \rangle\langle s,g|,
\end{equation}
where $H$ has $n$ different eigenvalues $\{E_s\}_{s=1}^n$, and each eigenvalue $E_s$ has $k_s$ degenerate eigenstates $\{|s,g\rangle\}_{g = 1}^{k_s}$. Here the index $s$ is for the $s$-th different eigenvalue, and the index $g$ is denoted as the $g$-th $E_s$ (the total number of the same eigenvalue $E_s$ is $k_s$). $|s,g\rangle$ is the eigenstate corresponding to the eigenvalue, i.e., the $g$-th $E_s$. According to the definition of block coherence, one can naturally choose the degenerate subspaces of $H$ in Eq. (\ref{1}) as the reference subspaces. Therefore,
\begin{equation}\label{Ps}
P_s = \sum_{g = 1}^{k_s} |s,g\rangle\langle s,g|,
\end{equation}
is the $s$-th subspace projector, where every pure state in this subspace is an eigenstate of $H$ with the eigenvalue being $E_s$.

\begin{figure}
\includegraphics[scale=1.3]{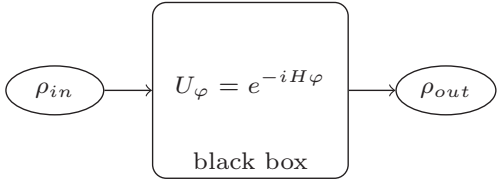}
\caption{The black box implements an unitary evolution $U_{\varphi}=e^{-iH\varphi}$ on the input state $\rho_{in}$, and the unknown parameter $\varphi$ of the black box must be estimated by measuring $\rho_{out}=U_{\varphi}\rho_{in}U_{\varphi}^{\dag}:=\rho_{\varphi}$. Suppose that the Hamiltonian $H$ is degenerate and already known, the quantum parameter estimation task is estimating the unknown parameter $\varphi$ from the output state $\rho_{out}$.}\label{Fig1}
\end{figure}

\textbf{Proposition 3.} With a degenerate Hamiltonian $H$ in Eq. (\ref{1}), we choose $\{P_s\}_{s=1}^n$ in (\ref{Ps}) as the reference subspace projectors $\mathrm{\mathbf{P}}$. The output state $\rho_{out}=U_{\varphi}\rho_{in}U_{\varphi}^{\dag}:=\rho_{\varphi}$ can be use to estimate the unknown parameter $\varphi$ in the black box in Fig. \ref{Fig1}, if and only if $\rho_{in}$ and $\rho_{out}$ have nonzero block coherence under the reference subspaces $\mathrm{\mathbf{P}}$.

\textbf{Proof.---} For an arbitrary input state $\rho_{in}$, it can be expressed under the eigenstates of $H$  (\ref{1}), i.e.,
\begin{equation}
\rho_{in} =\sum_{s,s'} \sum_{g,g'}\rho_{(s,g),(s',g')}|s,g\rangle\langle s',g'|,
\end{equation}
where we used $\rho_{(s,g),(s',g')} := \langle s,g|\rho_{in}|s',g'\rangle$ with $\sum_{s,g}\rho_{(s,g),(s,g)} = 1$. Thus,
the corresponding output state $\rho_{out} = U_{\varphi}\rho_{in} U_{\varphi}^{\dag}:=\rho_{\varphi}$ can be expressed as follows,
\begin{eqnarray}
\rho_{out}&=& U_{\varphi}\rho_{in} U_{\varphi}^{\dag}\nonumber\\
&=&\sum_{\substack{s,s'\\s\neq s'}}\sum_{g,g'}\rho_{(s,g),(s',g')}e^{-i(E_{s} - E_{s'})\varphi}|s,g\rangle\langle s',g'|\nonumber\\
&& + \sum_{s}\sum_{g,g'}\rho_{(s,g),(s,g')}|s,g\rangle\langle s,g'|.
\end{eqnarray}
We can see that $\rho_{out}$ is dependent on $\varphi$ if and only if there exists nonzero $\rho_{(s,g),(s',g')}$ with $s\neq s'$, i.e., $\rho_{out}$ and $\rho_{in}$ have nonzero block coherence under the reference subspaces. \hfill  $\square$

\textbf{Remark.---} It is worth noticing that witnesses for standard coherence may be useless to estimate the unknown parameter when the Hamiltonian $H$ is degenerate, although the input state has standard coherence. Consider a special case, if we use an input state which is block incoherent, but its density matrix contains off-diagonal nonzero elements in some degenerate subspaces under a chosen eigenvectors of  $H$ as reference basis, i.e., this input state contains standard coherence but no block coherence. The expect values of standard coherence witnesses has no information of the parameter, and one cannot estimate it by measuring witnesses of standard coherence, although the input state contains standard coherence.

\subsection{Quantum Fisher information of block incoherent states}
For an arbitrary output state $\rho_{\varphi}$, the quantum Fisher information $F_q$ can be obtained \cite{SCG,SCG2,MGA,VGSLM}
\begin{eqnarray}
F_q&=&\tr[\rho_{\varphi}L_{\varphi}^{2}]\nonumber\\
&=&\sum_{m,n}4c_{m}(\frac{c_{n}-c_{m}}{c_{n}+c_{m}})^2 |\langle m|H|n\rangle|^2,
\end{eqnarray}
where $U_{\varphi}=e^{-iH\varphi}$ is an unitary operator and $H$ is the corresponding Hermitian Hamiltonian, and the output state $\rho_{\varphi}=U_{\varphi}\rho_{in}U_{\varphi}^{\dagger}=U_{\varphi}(\sum_{n}c_{n}|n\rangle\langle n|)U_{\varphi}^{\dagger}$ with $\rho_{in}=\sum_{n}c_{n}|n\rangle\langle n|$.  $\{c_{n}\}$ and $\{|n\rangle\}$  are the eigenvalues and eigenvectors of $\rho_{in}$, respectively. The symmetric logarithmic derivative operator is $L_{\varphi}=U_{\varphi}(-2i\sum_{m,n}\frac{\langle m|[H,\rho_{in}]|n\rangle}{c_{n}+c_{m}}|m\rangle\langle n|)U_{\varphi}^{\dagger}$ \cite{SCG,SCG2,MGA,VGSLM}.
Furthermore, we will discuss the quantum Fisher information with a block incoherent state.

\textbf{Proposition 4.} Consider a degenerate Hamiltonian $H$ in Eq. (\ref{1}) and  the quantum parameter estimation task in Fig. \ref{Fig1}, one can choose the degenerate subspaces of $H$ as the reference subspaces. If the input state $\rho_{in}$  is a block incoherent state,  then the quantum Fisher information for the output state is $F_q=0$.

\textbf{Proof.---} Consider an arbitrary block incoherent state as the input state $\rho_{in}$,
\begin{eqnarray}
\rho_{in}&=&\sum_{i}\sum_{g,g'}\rho_{(i,g),(i,g')}|i,g\rangle\langle i,g'|\nonumber\\
&=&\sum_{i}\sum_{\tilde{g}}A_{\tilde{g}}^{(i)}|i,\tilde{g}\rangle\langle i,\tilde{g}|,
\end{eqnarray}
where the second equation holds since we diagonalize $\rho_{in}$ in each subspace. Thus, the eigenvalues of $\rho_{in}$ are $\{A_{\tilde{g}}^{(i)}\}$ with corresponding eigenvectors $\{|i,\tilde{g}\rangle\}$. The output state $\rho_{\varphi}$ can be expressed as
\begin{eqnarray}
\rho_{\varphi}&=&U_{\varphi}\rho_{in}U_{\varphi}^{\dagger}\nonumber\\
&=&U_{\varphi}\sum_{i}\sum_{\tilde{g}}A_{\tilde{g}}^{(i)}|i,\tilde{g}\rangle\langle i,\tilde{g}|U_{\varphi} ^{\dagger},
\end{eqnarray}
where $U_{\varphi}=e^{-i H\varphi}$ is an unitary operator. Furthermore, the symmetric logarithmic derivative operator $L_{\varphi}$ can be expressed as \cite{SCG,SCG2,MGA,VGSLM},
\begin{eqnarray}
L_{\varphi}
&=&U_{\varphi}(-2i\sum_{i,j}\sum_{\tilde{g},\tilde{h}}\frac{\langle i,\tilde{g}|[H,\rho_{in}]|j,\tilde{h}\rangle}{A_{\tilde{g}}^{i}+A_{\tilde{h}}^{(j)}}|
i,\tilde{g}\rangle\langle j,\tilde{h}|)U_{\varphi}^{\dagger}\nonumber\\
&=&U_{\varphi}\Bigg(-2i\sum_{i,j}\sum_{\tilde{g},\tilde{h}}\frac{A_{\tilde{h}}^{(j)}-
A_{\tilde{g}}^{(i)}}{A_{\tilde{g}}^{(i)}+A_{\tilde{h}}^{(j)}}\langle i,\tilde{g}|H|j,\tilde{h}\rangle |i,\tilde{g}\rangle\langle j,\tilde{h}|\Bigg)U_{\varphi}^{\dagger}.\nonumber
\end{eqnarray}
Finally, we obtain the quantum Fisher information $F_q$ as
\begin{eqnarray}
F_q&=&\tr(\rho_{\varphi}L_{\varphi}^{2})\nonumber\\
&=&4\sum_{i,j}\sum_{\tilde{g},\tilde{h}}A_{\tilde{g}}^{(i)}\big|\langle i,\tilde{g}|H|j,\tilde{h}\rangle\big|^{2}\Bigg(\frac{A_{\tilde{h}}^{(j)}-A_{\tilde{g}}^{(i)}}
{A_{\tilde{g}}^{(i)}+A_{\tilde{h}}^{(j)}}\Bigg)^2\nonumber\\
&=&4\sum_{\substack{i,j \\i\neq j}}\sum_{\tilde{g},\tilde{h}}A_{\tilde{g}}^{(i)}\big|\langle i,\tilde{g}|H|j,\tilde{h}\rangle\big|^{2}\Bigg(\frac{A_{\tilde{h}}^{(j)}-A_{\tilde{g}}^{(i)}}
{A_{\tilde{g}}^{(i)}+A_{\tilde{h}}^{(j)}}\Bigg)^2 \nonumber\\
&&+4\sum_{i}\sum_{\tilde{g},\tilde{h}}A_{\tilde{g}}^{(i)}\big|\langle i,\tilde{g}|H|i,\tilde{h}\rangle\big|^{2}\Bigg(\frac{A_{\tilde{h}}^{(i)}-A_{\tilde{g}}^{(i)}}
{A_{\tilde{g}}^{(i)}+A_{\tilde{h}}^{(i)}}\Bigg)^2 \nonumber\\
&=&0,\label{35}
\end{eqnarray}
where we have used $\langle i,\tilde{g}|H|j,\tilde{h}\rangle= 0$ with $i \neq j$, and $\langle i,\tilde{g}|H|i,\tilde{h}\rangle=E_i \delta_{\tilde{g}\tilde{h}}$. \hfill  $\square$

\textbf{Remark.---} When the input state is a block incoherent state, the quantum Fisher information of the output state is equal to zero, which means one cannot estimate the parameter from the output state, i.e., the output state contains no information of the parameter. That coincides with the result from Proposition 3.

\section{Discussions and conclusions}
Besides the Fisher information estimation mentioned in the above example, when performing quantum coherence operations on any degenerate quantum system, we also need to take into account the measure of coherence. In the practical systems, degeneracy is a very common situation and has been widely reported in the microscopic field, e.g., orbital degenerate states and degenerate spin states in $\lambda$-type and the chainlike structures atomic system \cite{EXP1}, near-degenerate states in few-electron ions system \cite{EXP2}, etc. Moreover, degeneracy also exists in macroscopic systems, e.g., macroscopically degenerate in intrinsic quasi-crystals \cite{EXP3}, and been regarded to be appeared during the conversion or transition of black holes \cite{EXP4}. In addition, degeneracy plays an important role in both quantum computing and quantum networks construction \cite{EXP5,EXP6}. It is often used as a protected qubit in fault-tolerant quantum operation \cite{EXP5}. And it can also be used as a control device to control the coherence between quantum systems \cite{EXP7}. In the discussion of topological structure and phase transition, the study of degeneracy and near-degeneracy also plays an important role \cite{EXP8,EXP9,EXP10}. These studies related to the issue of how to correctly measure the coherence of degenerate quantum systems. Therefore, our coherence detection scheme has potential value in the measurement and application of degenerate states.

In conclusion,  we have discussed witnesses for block coherence and POVM-based coherence. The necessary and sufficient conditions for arbitrary block coherence  and POVM-based coherence witnesses have been obtained. Furthermore, we have shown that block coherent states can be used in a quantum parameter estimation task with a degenerate Hamiltonian if the input state is block coherent. The quantum Fisher information of block incoherent states is equal to zero, which coincides with the result from the block coherence witness.

\section*{ACKNOWLEDGMENTS}
This work is supported by the National Natural Science Foundation of China (Grants No. 11734015, No. 11704205, and No. 12074206), the open funding program from State Key Laboratory of Precision Spectroscopy (East China Normal University), and K.C. Wong Magna Fund in Ningbo University.

\end{document}